\documentclass[10pt, conference, compsocconf]{IEEEtran}
\ifCLASSINFOpdf
\else
\fi
\usepackage{mathtools}
\usepackage{cases}
\usepackage{graphicx}
\usepackage{amsmath}

\begin{document}

\title{Analytical Survey of Wearable Sensors}
\author{A. Rehman, M. Mustafa, N. Javaid, U. Qasim$^{\ddag}$, Z. A. Khan$^{\S}$\\

        COMSATS Institute of IT, Islamabad, Pakistan. \\
        $^{\ddag}$University of Alberta, Alberta, Canada\\
        $^{\S}$Faculty of Engineering, Dalhousie University, Halifax, Canada.
        }
\maketitle

\begin{abstract}
Wearable sensors in Wireless Body Area Networks (WBANs) provide health and physical activity monitoring. Modern communication systems have extended this monitoring remotely. In this survey, various types of wearable sensors discussed, their medical applications like ECG, EEG, blood pressure, detection of blood glucose level, pulse rate, respiration rate and non medical applications like daily exercise monitoring and motion detection of different body parts. Different types of noise removing filters also discussed at the end that are helpful in to remove noise from ECG signals. Main purpose of this survey is to provide a platform for researchers in wearable sensors for WBANs.

\end{abstract}

\begin{IEEEkeywords}
Wearable Sensors; Accelerometers; ECG; Physical Activity
\end{IEEEkeywords}
\IEEEpeerreviewmaketitle

\section{Introduction}

Rapid increase in world population of elderly people have drawn attention from researchers to develop a system that reduces health-care cost, efficient utilization of physician skills, remote access to patients for continuous monitoring and analysis feedback to patients to reduce severe health related issues. Wireless wearable sensors are major part of this health-care system, that works as sensing node and measure different physiological signals such as heart rate, body and skin temperature, blood pressure, Electrocardiography (ECG), Electroencephalogram (EEG), Electromyography (EMG) signals, oxygen saturation and respiration rate etc. These collected signals transferred to a central node via wireless or wired medium. For further processing and analysis of disease, these signals transmitted to medical server via wireless medium.

Mobility is a key part in health-care system, for this purpose wearable sensors must be small in size, power efficient, low weight and should have wireless module for wireless communication.


\section{Types of Wearable Sensors}
With increase in population and changing life styles there is urgency to develop a system that can monitor patient activities and daily routines to prevent them from serious health related disorders \cite{yilmaz2010detecting}. Advancements in wearable sensors and wireless technologies create huge impact on health-care monitoring system. Now we have facilities to monitor patients from remote location on continuous basis by using wearable sensors and wireless systems. Different types of sensors available for specific applications.

\textit{A. Accelerometer: }
Accelerometer sensors or motion detection sensors are used to sense acceleration (change in body position), this acceleration might be linear or angular. Operational principle of accelerometer is based on an element named proof mass that attached to a suspension system with respect to reference point and when force applied on proof mass, deflection is produced in it. Produced deflection can be measured electrically to sense changes in body location \cite{godfrey2008direct}. Accelerometers are most commonly used sensors to monitor physical activities of persons who recently recovered from brain disease \cite{casson2008wearable}. It specifically used in rehabilitation process of stroke and parkinson survivors to check the level of mobility, also used in analysis of gait.

\textit{B. Electromagnetic Tracking System (ETS) Sensor: }
ETS is a body position measurement sensor based on Faraday’s law of magnetic induction \cite{raab1979magnetic}. When a person or object that carry a sensor consists of coils perform a motion inside a controlled magnetic field, the induced voltage in sensor coils will change with respect to the change of the objects position and orientation relative to source of controlled magnetic field. This controlled magnetic field is generated by a fixed transmitter and detected by a receiver fixed on an object. By using this phenomena position and orientation of moving object can be calculated \cite{tao2012gait}. ETS is an important sensor in gait analysis and in study of body kinematics.

\textit{C. Ground Reflection Force (GRF) Sensor: }
GRF sensor is used to realize ambulatory measurements of ground reflection force during gait analysis. It is a three dimensional vector, with actual direction depending upon the nature of interface between ground and foot. Shoe based GRF sensor is an alternative of old conventional techniques that were used in laboratory for gait analysis such as instrumented treadmill devices \cite{tao2012gait}. In \cite{liedtke2007evaluation}, authors developed a shoe based GRF sensor by fixing two externally mounted sensors beneath front and rear part of a special shoe. In \cite{tao2012gait}, authors proposed a new shoe based GRF sensor by using five small triaxial sensors beneath shoe. They aligned each coordinate of sensor with global coordinate systems; then collect data about each sensor position in accordance to reference positions and use this data to analyze different parameters. This GRF sensor used to measure Center of Pressure (CoP) in ambulatory measurements and also used to analyze kinetics of ankle, knee and hip joints.

\textit{D. EMG Sensor: }
In EMG electrical activities of particular muscle is monitored. During muscle contraction microvolt level electrical signals produced, that can be measured from skin surface. In other words EMG measures the action of muscles. Basically two types of EMG sensors are used, needle EMG and surface EMG. Surface EMG or sEMG is used when only basic or general information of muscle activity is required, whereas, in needle EMG, needle must be inserted inside designated muscle which required to be studied. Needle EMG sensors are used to acquire some detailed information about specific muscle \cite{davis1988clinical}. EMG specifically used to study the performance of persons who suffered from skeletal problems for example used in localized muscle fatigue and gait analysis to study muscle force.

\textit{E. ECG Sensor: }
ECG is interpretation of electrical activity of heart over a period of time across chest area whose purpose is to record activities of heart during its contraction and relaxation. In conventional methods electrodes were attached on body surface around chest that measures electrical signals during heart contraction process. Received signals from electrodes were recorded to an external device called holter. It is impossible from traditional system to perform ECG at remote location. With the advancements in technology different ideas were presented to replace wired holter with wireless holter system. Design of electrodes is also important factor in continuous monitoring that these electrodes should not damage the skin. Different electrodes were used to monitor heart activities from remote locations for continuous period, for example use of dry electrodes, electrodes made up of plastic or rubber. However these type of electrodes cause skin irritation problems.

In \cite{chi2010wireless}, authors proposed an idea to use non-contact capacitive sensing mechanism, in which capacitive electrodes can sense heart signals through clothes. They propose an idea of using two gold coated electrodes on each arm (wrist) surface and record ECG by using single channel between each arm and results show an error heart rate within range of 1\%. In \cite{chung2008wireless}, authors develop a single chip based ECG sensor that consists of two conductive fabric electrodes to detect heart signals. This wearable ECG sensor amplifies detected signals and then transmits to server.

\textit{F. EEG Sensor: }
EEG is a process to measure brain waves of a person, in its conventional method a number of electrodes are placed on scalp; these electrodes detect microvolt level signals coming from brain. Currently different methods been adopted to measure EEG for example Inpatient and Ambulatory EEG methods. But these methods also have some limitations like mobility. In Inpatient EEG method a person have to present in hospital for EEG and in Ambulatory EEG (AEEG) method a person can perform EEG at anywhere but it also has a limited mobility level because EEG monitoring system have box like device that a person have to carry all the time, and this is not a desirable situation for anybody. To overcome these issues a number of researchers present ideas about Wearable EEG sensors.

In \cite{casson2008wearable}, authors conduct a survey about adoptability of Wearable EEG sensors in future and they got a very good response about it. After this they propose a novel design approach of wearable EEG. In first approach they propose that wearable system of electrodes should be wireless to get rid of electrode wires, in second approach they give an idea to use Dry electrodes instead of wet or gel based electrodes. These two approaches have a drawback of placement of electrodes on scalp for long duration. To overcome this, they provide a solution to place electrodes beneath scalp skin. This approach has several advantages like electrodes will remain invisible; they will not further misplace and can be used to monitor EEG for up to eighteen months. EEG sensors are specifically used in Epilepsy and sleep studies.

\textit{G. Blood Glucose Monitoring Sensor: }
In conventional methods of Blood Glucose (BG) monitoring, blood sample is obtained from body by placing blood sample on a strip and then insert it into a BG calculating device to calculate Blood Glucose Level (BGL). However this conventional method is based on invasive technique, not suitable for continuous monitoring. A commercial wearable BGL monitoring sensor was developed, that has minimal invasive effect. A needle consists of electronic chip is inserted into human body to take blood sample, process it and send results wirelessly to server system. But due to shorter life duration a lot of work required on this system. Some other invasive methods were also proposed that used for continuous monitoring, these methods based on the concept of extracting fluid from skin with the help of some vacuum pressure to measure BGL \cite{yilmaz2010detecting}.

Some other methods of measuring BGL non-invasively were also presented, for example by checking electrical properties of blood we can estimate BGL. But in non-invasive methods a lot of work required to be done.

\section{Design of Wearable Sensors}

 For efficient utilization of physician’s resources and health related cost, researchers and experts propose the idea of ubiquitous health care system. Ubiquitous health care systems provide a smarter and cheaper way to efficiently deal with patients suffering from chronic diseases \cite{schepps2002microwave}. For implementation of this system wireless wearable or implantable sensors required to monitor patient activities. Currently researcher’s main focus is to develop such sensors that are comfortable and non-invasive, utilize minimum energy and provide maximum and accurate results \cite{lubrin2005wireless}. In following sections a brief survey of wearable sensors will be given regarding their design.

\textit{A. Non-contact EEG/ECG Sensor Electrode: }
EEG and ECG signals from brain and heart are most critical parameters to be monitored in long term continuous health monitoring system by using wearable sensors. Conventionally wet EEG and ECG electrodes were used for monitoring signals, after some technological advancements use of dry electrodes instead of wet become common, but due to their continuous use some skin related problems arise. After this researchers divert their focus to develop minimal or non-invasive technique to measure these critical health parameters.
In \cite{chi2010wireless}, authors develop a Non-contact EEG/ECG wireless sensor electrode to detect signals from brain and heart. Upper Printed Circuit Board (PCB) contains a low noise amplifier and 16 bit Analog to Digital Converter (ADC) that output detected signals in digitized values. Whereas, lower PCB consists of amplifier (INA116), bottom surface of PCB filled with solid copper and insulated by soldermask, that works as electrodes to detect signals from surface.

In development of non-contact low noise electrode sensor, main challenge is to design an ultra-high input impedance and low noise amplifier. For this purpose authors of \cite{chi2010wireless}, design a circuit for electrode sensor. It consists of voltage source $V_{s}$ that is connected to input of amplifier, who has coupling capacitance $C_{s}$ with finite resistance $R_{b}$ and input capacitance $C_{in}$. This amplifier has a positive feedback that is applied through $C_{n}$. Input voltage noise of amplifier is $V_{na}$, input current noise is $I_{na}$, whereas, additional current noise is given by $I_{nb}$. Total input noise of capacitive amplifier is given by this equation.
\begin{eqnarray}
  V_{n}^{2} =V_{na}^{2}(1+\frac{c_{in}+c_{n}}{c_{s}})^{2}+\frac{i_{na}^{2}+i_{nb}^{2}}{W^{2}C_{s}^{2}}
\end{eqnarray}

As physiological signals have very low frequencies and even a small amount of current noise cause huge input voltage noise. Authors use bias free technique to match noise specifications of amplifier (INA116). However cut-off frequency was set to 0.7Hz with a gain of 2.02dB. To operate this electrode over different coupling distances they use positive feedback technique. Output from amplifier (INA116) is forwarded to another amplifier (LTC6078) passed through a high-pass amplifier having cut-off frequency of 0.1Hz with 40.01 dB gain. These electrodes connected to wireless base unit, that receives all data from electrodes and forward it to monitoring server. In this system possibility of getting extra noise from external sources is a problem.

\textit{B. PTT based Blood Pressure Estimation: }
Pulse Transit Time (PTT) is a method use to estimate blood pressure non-invasively. PTT measure the time taken by a pulse wave to travel between two points in circulatory system. Pulse Wave Velocity (PWV) calculated by using following equation.
\begin{equation}
C^{2}=\frac{\Delta pV}{\Delta v\rho}
\end{equation}
where, $C$ denotes PWV, $\Delta p$ is change in pressure, V is the initial volume, $\Delta v$ shows change in volume and $\rho$ is density of fluid. PTT can be calculated as
\begin{equation}
PPT=\frac{1}{PWV}
\end{equation}
\begin{figure}[h!]
    \centering
        \includegraphics[width=9cm,height=4.5cm]{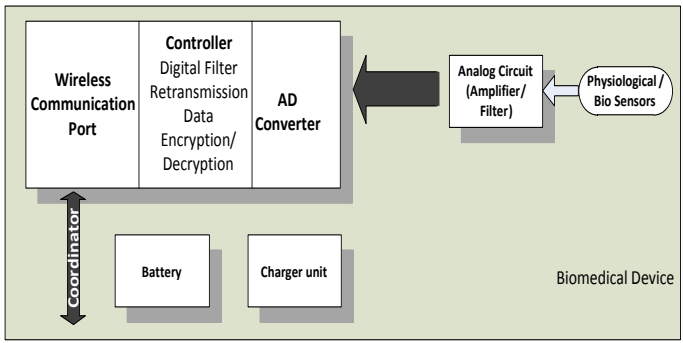}
    \vspace{-0.7cm}
    \caption{Functional Block Diagram}
\end{figure}
In \cite{kim2010cuffless}, authors develop a wireless wearable device that consists of several bio signal measuring modules. It has measuring sensors for ECG, Photoplethysmograph (PPG) that measures changes occur in blood optically, skin surface temperature, fall detection and Non-Invasive Systolic Blood Pressure (NISBP). Micro-controller works as a central processor, which manage all operations of attached sensors. ECG sensor has two electrodes used to detect heart signals at two different locations on wrist. A flexible ribbon type sensor used to measure skin temperature. SpO2 sensor is attached on top of wrist band such that fingers of other hand easily touch on its surface to detect PPG signals from finger. This device consists of micro-controller, signal detecting sensors, analog circuits, ADCs and wireless modules.

To measure ECG with the help of wrist worn device two electrodes adjusted in such a way that one electrode must sense signals from wrist on which patient wear this device, second electrode place on top of device such that other hand can easily touch surface of electrode. This ECG module consists of instrumentation amplifier, notch filer, and non-inverting amplifier with bandwidth of 50Hz. Detected signals is then digitized for transmission and evaluation.

To measure skin surface temperature, a flexible ribbon type sensor used, that is attached with inner surface of device such that sensor can touch patient skin to measure temperature. Whereas, fall detector sensor is a 3-axis accelerometer, when cumulative value of all axis reaches a threshold a fall event occurs. An increase in blood pressure increases PWV \cite{wei1990optimal}, by detecting this effect with the help of ECG and PPG systolic blood pressure can be measured.

\textit{C. Cuff-less PPG Based Blood Pressure Monitoring: }
Cuff-based oscillometric devices used for continuous ambulatory blood pressure monitoring. To estimate blood pressure, relationship of external pressure with magnitude of arterial volume pulsation is used. However this traditional method is not suitable for long term monitoring. In \cite{shaltis2006wearable} authors develop a PPG based non-invasive continuous blood pressure monitoring method. PPG uses optical signals to measure volumetric pulsation of blood in tissues. This wearable device has some technical issues that must be noted. Measurement of Mean Arterial Pressure (MAP) requires an effective method to check volumetric changes in blood. To measure hydrostatic pressure offset against heart, a height sensor is required that should be wearable, compact in size and consume low power. Following equation is used to measure pressure difference across vascular wall.
\begin{equation}
P_{tm} = P_{MAP} -\rho.g.h-P_{cuff}
\end{equation}
where, $P_{tm}$ is Transmural Pressure, $P_{MAP}$ is Mean Arterial Pressure, $\rho.g.h$ is pressure offset when location of measuring device is not as same as heart however this value will be omitted from equation if height of measuring device and heart have same height levels and $P_{cuff}$ is pressure applied from external source. A known amount of pressure (below 75mmHg) is applied from cuff based device and when it matches with internal MAP, a large amplitude pulse is detected (Zero Transmural Pressure point). PPG is used to detect changes in volume of blood vessels. To overcome the problem of applying large pressure across cuff, authors use concept of raise and lower arm to alter pressure in vessels.

Authors define following procedure to measure blood pressure. By fixing pressure across cuff, PPG sensor worn arm is raised to check variations in reference pressure.
\begin{equation}
P_{r} = \rho.g.h+P_{cuff}
\end{equation}
and PPG signal having highest amplitude shows zero transmural pressure point.
\begin{equation}
P_{MAP} = P_{r} = \rho.g.h+P_{cuff}
\end{equation}

To precise motion of PPG worn arm, authors introduce accelerometer based arm movement control process. They attach two accelerometer sensors on arm; first accelerometer is attached on bicep area and second is on finger (embedded with PPG). Where, $l_{o}$ is distance from shoulder to heart, $l_1$ length of upper arm and $l_2$ is length of forearm.

Following equations use to measure height of PPG sensor with respect to heart.
\begin{equation}
h = l_{1}.\cos\theta_{1} + l_{2}.\cos\theta_{2}-l_{o}
\end{equation}
\begin{equation}
h = l_{1}.\sin\theta_{1} + l_{2}.\sin\theta_{2}-l_{o}
\end{equation}

\textit{D. sEMG Electrode based Sensor: }
In \cite{al2011autonomous}, authors design a sEMG electrode based sensor to check properties of bicep muscle with the help of goniometer sensor. Developed system consists of two parts. Amplification part contains an amplifier and filtering circuit, and a SunSPOT that contains different circuitry for processing of received signals from bicep worn sEMG sensor.

Signals coming from body surface have very small peak to peak amplitude, to amplify these signals; amplifier is directly connected with leads coming from body surface. Received signals amplify about 330 times of original signals and filtered with 10 to 1000 Hz bandpass filter.

SunSPOT is a processing unit produced by Sun Microsystems that consists of a microcontroller and 10 bit ADC. Analogue signals coming from bandpass filter forwarded to ADC for conversion, and these converted signals send to server for processing via wireless medium.
\begin{figure}[h!]
    \centering
        \includegraphics[width=7.5cm,height=3.7cm]{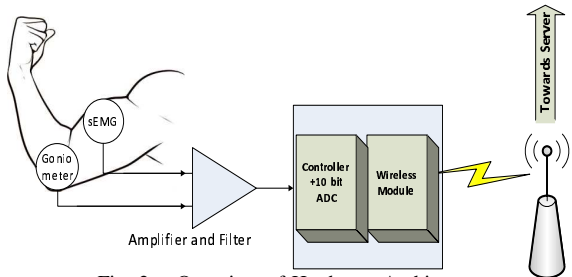}
    \vspace{-0.5cm}
    \caption{Overview of Hardware Architecture}
\end{figure}
\section{ECG Noise Removing Filters}

In health-care monitoring system, wearable sensors measure different types of physiological signals, like ECG, EEG, EMG etc. After passing through different devices and mediums, these signals contain different types of noises. For analysis of these signals, they must be in noise free form. For processing of these signals, a process or device named Filter is used to remove unwanted noise. Normally filters are used to suppress aspects of signals completely or partially depending upon noise to be removed. However while filtering these signals, filters might remove required information associated with noise \cite{mashaghi2011noise}.

In digital signal processing applications, digital filters are most important elements. These digital filters might have been categorized as Finite Impulse Response (FIR) and Infinite Impulse Response (IIR) filters with respect to their duration of impulse responses.

\textit{A. FIR Filters: }
FIR filters are widely used due to their powerful design, inherent stability and linear phase. These filters have impulse response of finite durations, after this finite duration it settles to zero.
\begin{equation}
y[n]=b_{0}x[n]+b_{1}x[n-1]+...+b_{k}x[n-k]
\end{equation}
\begin{equation}
y[n]=\sum_{k=0}^{M}b_{k}x[n-k]
\end{equation}
where, $x[n]$ is input signal, $y[n]$ is output signal, $b_{i}$ is filter coefficients and $N$ is the filter order. These filters output is only dependent upon present and previous values of input. However these filters have high complexity issues. FIR filter can be further classified into two categories: Window based and Frequency sampling domain methods. However, only window based methods will be discussed here.

\textit{1. Kaiser Window: }
The Kaiser window is an approximation to a restricted time duration function with minimum energy outside some specified band. If we have information about ripples and transition bandwidth then by using following equations we can find remaining parameters.
\begin{equation}
\alpha=-20\log_{10} \text{(Amount of Ripples Allowed)}
\end{equation}
where, $\alpha$ is side lobe attenuation in dB. Width of smallest transition region can be calculated by using this equation.
\begin{equation}
\Delta\omega=2\pi \frac{\text {Transition Width}}{\text{Sampling Frequency}}
\end{equation}
Now for filter order following equation is used
\begin{equation}
 N =
  \begin{dcases}
   \frac{\alpha-7.95}{2.285\Delta \omega} & \text{if } \alpha > 21\\
  \frac{5.79}{\Delta\omega} & \text{if } \alpha \leq 21
  \end{dcases}
\end{equation}
\begin{equation}
 \beta =
  \begin{dcases}
   0.1102(\alpha-8.7) & \text{if } \alpha > 50\\
   0.582(\alpha-21)^{0.4}+0.07887(\alpha-21) & \text{if} 21 \leq \alpha \leq 50\\
   0 & \text{if } \alpha < 21
  \end{dcases}
\end{equation}
where, $\beta$ is parameter that affects the side lobe attenuation, increasing beta widens main lobe due to this attenuation will increase.

\textit{2. Hanning Window: }
The Hann or Hanning window, belongs to family named "raised cosine" windows, the term "Hanning window" is sometimes used to refer to Hann window. Coefficients of a Hanning window can be computed from following equation.
\begin{align}
  \omega(n)=0.5
  \left(
    1-\cos\left(2\pi \frac{n}{N}\right)
  \right)
\end{align}
where, $N$ is order of window.

\textit{3. Hamming Window: }
The "raised cosine" with these particular coefficients was proposed by Richard W. Hamming. Coefficients of a Hamming window can be computed from following equation.
\begin{equation}
  \omega(n)=0.54-0.46\cos(2\pi\frac{n}{N}),  0 \leq n \geq N
\end{equation}
where, $N$ is order of window.

\textit{4. Blackman Window: }
In Blackman window side lobes rolloff at about $18dB$ per octave. Coefficients of Blackman window are calculated as
\begin{align}
\omega(n)& = 0.42-0.5\cos(\frac{2\pi n}{2N+1})   \nonumber \\ & \qquad {}
 +0.08\cos(\frac{4\pi n}{2N+1}), -N \leq n \geq N
\end{align}
Number of terms for Balackman window is give as
\begin{equation}
N^{'}= 5.98\frac{f_{s}}{T.W}
\end{equation}
$f_{s}$ is sampling frequency and $T.W$ is transition width.

\textit{5. Blackman-Harris Window: }
Blackman-Harris (BH) window family is generalization of Hamming family. Coefficients of BH window are calculated as
\begin{align}
\omega(n)& =  0.358 + 0.488\cos(\frac{2\pi n}{N+1})   \nonumber \\ & \qquad {}
 + 0.142\cos(\frac{2\pi n}{N+1}) + 0.012\cos(\frac{2\pi n}{N+1})
\end{align}
where, $-\frac{N}{2}\leq n \leq \frac{N}{2}$

\textit{B. IIR Filters: }
Digital filters which must be implemented recursively are called Infinite Impulse Response (IIR) filters because, theoretically, the response of these filters' to an impulse never settles to zero. IIR filters output completely depend upon previous inputs, present inputs and on previous outputs. These filters are very helpful for designing high speed signal processing, because these types of filters have less number of multiplications as compared to FIR filters. Difference equation or response of filter is given by following equation.
\begin{equation}
y[n]=-\sum_{k=1}^{N}a_{k}y[n-N]+\sum_{k=1}^{M}b_{k}x[n-M]
\end{equation}
where, $N$ is feedforward filter order, $M$ is feedback filter order, $a_{i}$ feedforward coefficient, $b_{i}$ is feedback coefficient, $x[n]$ is input signal and $y[n]$ is output signal. First part of equation represents recursive part of IIR filter and second part shows non-recursive part. Different types of IIR filters will be discussed briefly here.

\textit{1. Butterworth Filter: }
Butterworth filters are characterized by a magnitude response that is maximally flat in the passband and monotonic overall. Decay is slow in passband and fast in stopband, due to this, it is preferable choice where low number of ripples required in pass and stopband.
\begin{align}
  |H(\omega)|=\left[\frac{1}
    {1+\left(\frac{\omega}{\omega{c}}\right)^{2N}}
  \right]^{\frac{1}{2}}
\end{align}
If we have values of pass and stopband attenuations and frequencies then by using this equation value of cut-off frequency and order of filter can calculated. Values of cut-off frequency and order of filter then further used to calculate the filter transfer function.

\textit{2. Chebyshev I filter: }
In Chebyshev Type I faster roll-off can be acquired by allowing ripple in the frequency response. Analog and digital filters that use this approach are called Chebyshev filters. These filters are named from their use of Chebyshev polynomials, developed by Russian mathematician Pafnuti Chebyshev.
Chebyshev Type I filter has magnitude response given by following equation
\begin{align}
  |H(\omega)|=\frac{A}{\left[
    1+\varepsilon^{2}C_{N}^{2}\left(\frac{\omega}{\omega_{c}}\right)
  \right]^{\frac{1}{2}}}
\end{align}
where, $A$ is filter gain, $\omega_{c}$ is cut-off frequency, $\varepsilon$ is a constant, and filter order calculated using following equation
\begin{equation}
C_{N}(x)=\cos(N\cos^{-1} x)),  for   (x)\leq 1
\end{equation}
\begin{equation}
C_{N}(x)=\cos(N\cosh^{-1} x)),  for   (x)\geq 1
\end{equation}
\textit{3. Chebyshev II Filter: }
It is also known as inverse of chebyshev filter. Chebyshev Type II filters have ripple only in the stopband and it does not roll-off as fast as chebyshev Type I. Type II filters are seldom used.
\begin{align}
  |H(\omega)|=\frac{\varepsilon C_{N}(\frac{\omega_{c}}{\omega})}{\left[
    1+\varepsilon^{2}C_{N}^{2}\left(\frac{\omega_{c}}{\omega}\right)
  \right]^{\frac{1}{2}}}
\end{align}
where, $\varepsilon$ is a constant and $\omega_{c}$ is 3dB cut-off frequency.

\textit{4. Elliptical Filter: }
Elliptic or Cauer filters exhibit equiripple behavior in both passband and stopband. This type of filter contains both poles and zeros and is characterized by magnitude response
\begin{align}
  |H(\omega)|=\left[\frac{1}
    {1+\varepsilon^{2}U_{N}\left(\frac{\omega}{\omega{c}}\right)^{2N}}
  \right]^{\frac{1}{2}}
\end{align}
$U_{N}(x)$ is Jacobian elliptical function of order N, and $\varepsilon$ is a parameter related to the passband ripple. The order of elliptic filter that is required to achieve given specifications is lower than order of Chebyshev and Butterworth filters. Therefore elliptical filters form an important class, but the design of this filter is more complex than other filters.
\section{Conclusion and Future Work}
In this paper Wireless Wearable sensors has been discussed with respect to different motion detection scenarios and a brief survey of wireless wearable sensor designs. At the end we discussed Lowpass, Highpass and Notch filters for both IIR and FIR (windowing techniques) that are helpful to remove noise from raw ECG signals. Different techniques have been discussed to remove noise from physiological signal, implementation of these filters to remove different types of noises from raw ECG signal and analysis of these filters has been left for the future work.
\bibliographystyle{ieeetr}
\bibliography{bare_conf}
\end{document}